\documentclass[12pt]{article}

\begin{document}
%New environment
\newenvironment{Eqnarray}%
         {\arraycolsep 0.14em\begin{eqnarray}}{\end{eqnarray}}

%%%%titipage 
\begin{flushright} 
December, 1999 \\
KEK-TH-668 \\
TIT/HEP-439 \\
%{\tt hep-th/99nnnnn}  
\end{flushright} 

\vspace{1cm}

\begin{Large}
       \vspace{2cm}
  \begin{center}
   {IIB Matrix Model with D1-D5 Backgrounds}      \\
  \end{center}
    \end{Large}

  \vspace{1cm}

\begin{center}
           YUSUKE KIMURA $^{a)}$$^{b)}$\footnote{
e-mail address : kimuray@ccthmail.kek.jp or kimura@th.phys.titech.ac.jp}
and YOSHIHISA KITAZAWA $^{a)}$\footnote{
e-mail address : kitazawa@post.kek.jp} \\
        
       \vspace{0.7cm} 
        $a)${\it  High Energy Accelerator Research Organization (KEK),}\\
               {\it Tsukuba, Ibaraki 305-0801, Japan} \\
       \vspace{0.3cm}   
        $b)$ {\it  Department of Physics, Tokyo Institute of Technology, \\
 Oh-okayama, Meguro-ku, Tokyo 152, Japan}\\
\end{center}
        \vspace{0.8cm}

\begin{abstract}
\noindent
\end{abstract}

We consider IIB matrix model with 
D1-D5-brane backgrounds. 
Using the fact 
that noncommutative gauge theory on the D-branes 
can be obtained as twisted reduced model in IIB matrix model, 
we study two-dimensional gauge theory 
on D1-branes and D5-branes.
Especially the spectrum of the zero modes in the 
off-diagonal parts is examined. 
We also consider the description of D1-branes 
as local excitations of gauge theory on D5-branes. 
Relation to supergravity solution 
is also discussed.

\newpage

%%%%%%%%%%%%%%%%%%%%%%%%%%%%%%%%%%%%

\section{Introduction}
\hspace{0.4cm}
Several kinds of Matrix Model 
have been proposed\cite{BFSS,IKKT,DVV,IT} 
to study the nonperturbative aspects of string theory or M theory.
These proposals are based on the developments of D-branes.
They have been shown to play a fundamental role 
in nonperturbative string theory\cite{Dbrane,pol}.
Notable point is that 
supersymmetric gauge theory can be obtained 
on their world-volume 
as their low energy effective theory.
The idea of matrix models is that
supersymmetric gauge theory can 
describe string or M theory.

IIB Matrix Model is one of these proposals\cite{IKKT}.
It is  a large $N$ reduced model\cite{reduced} 
of ten-dimensional supersymmetric 
Yang-Mills theory. 
It is postulated that it gives the constructive definition of 
type IIB superstring theory. 
%This is based on the idea that 
%a large $N$ gauge theory is equivalent 
%to its reduced model\cite{reduced}.
It is crucial that this model 
has ${\cal N}=2$ supersymmetry 
because it guarantees the existence of gravitons.

Recently, noncommutative Yang-Mills theory has been 
studied in many situations.
It first appeared within the framework 
of toroidal compactification 
of Matrix theory\cite{CDS}. 
The world volume theory on D-branes 
with NS-NS two-form background 
is described by noncommutative Yang-Mills theory\cite{SW}. 
It was shown\cite{AIIKKT} that in matrix model picture 
noncommutative Yang-Mills theory 
is equivalent to twisted reduced model\cite{GAO}. 
In IIB matrix model, twisted reduced model is obtained as 
expanding the model around noncommutative backgrounds. 
A noncommutative background is a D-brane-like background 
which is a solution of equation of motion and 
preserves a part of supersymmetry. 
It is well known that gauge theory is realized 
in the world-volume of D-branes as their low energy effective theory. 
In IIB matrix model, 
gauge theory is realized as twisted reduced model. 
The remarkable point is that there is a relation 
between the coordinate space and the momentum space. 
%If we consider the theory in the low energy region compared 
%to the noncommutative scale, 
%we can ignore the noncommutativity.
Further researches 
about noncommutative Yang-Mills 
were done in \cite{IIKK}
using the property of twisted reduced model.
The behaviors of Wilson loop operators in 
the large momentum scale region 
and the small momentum scale region are studied. 
In small momentum scale region 
Wilson loops are the ones 
in ordinary gauge theory 
while in large momentum scale region 
Wilson loops  become open string like objects. 
It was pointed out that there was a crossover 
at noncommutative scale. 

In this paper, we study IIB matrix model with D1-D5 backgrounds.
The D1-D5 background is an interesting configuration 
since it is used by Strominger and Vafa\cite{SV} 
in order to study the microscopic interpretation 
of the black hole entropy. 
It is further studied by Maldacena\cite{Maldacena} 
as $AdS/CFT$ correspondence. 
The organization of this paper is as follows. 
In section 2, we review the relation between 
noncommutative Yang-Mills and IIB matrix model. 
Noncommutative Yang-Mills is equivalent to 
twisted reduced model.
Twisted reduced model is defined by the expansion 
around a D-brane-like background in IIB matrix model.
In section 3, we consider D1-D5 backgrounds. 
For their description, we divide matrices into four parts.
The upper left part and the lower right part 
represent D1-branes and D5-branes respectively.
The remaining parts represent the interactions of them. 
In string theory picture, 
off-diagonal parts represent the states 
appearing as open strings connecting D1-branes and 
D5-branes.
We examine the spectrum of this matrix model. 
The spectrum of open strings connecting D1-branes and 
D5-branes 
in string theory is discussed in \cite{SW}. 
Comparison with string theory results 
is mentioned. 
We also consider another description of 
D1-branes as local excitations of gauge theory 
on the world-volume of D5-branes.
This description gives the description of 
D1-branes bounded to D5-branes. 
The counting of the zero modes is also discussed. 
In section 4, we mention the relation to 
supergravity solutions. 
The relation between supergravity and gauge theory is 
formulated as $AdS/CFT$ 
correspondence\cite{Maldacena}.
B-field background effect changes the behaviors 
of gauge theory, 
while supergravity solutions are also changed\cite{HI,MR}.
Interactions in very far region from branes, flat spacetime, 
are described by the 
cluster property in matrix model.
In near horizon region (so called $AdS$ region) 
gauge theory description can be used.
We consider 
$AdS_3 \times S^3 \times M^{4}$ background in matrix model.
$AdS_5 \times S^5$ background is investigated in \cite{AIIKKT,IIKK}.
Section 5 is devoted to conclusions and discussions.

%%%%%%%%%%%%%%%%%%%%%%%%%%%%%%%%%%

\section{Noncommutative Yang-Mills in IIB matrix model}

\hspace{0.4cm}
In this section, we review 
IIB matrix model\cite{IKKT,AIKKTT} 
and its relation to noncommutative Yang-Mills 
which is equal to twisted reduced model\cite{AIIKKT,IIKK}.

IIB matrix model is defined by the following action
\begin{equation}
 S= -\frac{1}{g^{2}} Tr \left( \frac{1}{4} \left[ A_{\mu} ,A_{\nu}\right] 
         \left[A^{\mu} ,A^{\nu} \right]  
     + \frac{1}{2}\bar{\psi } \Gamma^{\mu} \left[ A_{\mu},\psi \right] \right) .         \label{action}
\end{equation}
Here $\psi$ is a ten dimensional Majorana-Weyl spinor field, 
and $A_{\mu}$ and $\psi$ are $N \times N$ hermitian matrices. 
This model is the large $N$ reduced model of ten-dimensional 
$\cal{N}$=1 supersymmetric Yang-Mills theory. 
This is based on the observation that 
a large $N$ gauge theory is equivalent 
to its reduced model\cite{reduced}.
This model has the following symmetries:
\begin{Eqnarray}
\delta^{(1)} \psi &=& \frac{i}{2} 
  \left[ A_{\mu} ,A_{\nu}\right] \Gamma^{\mu\nu} \epsilon  \label{susy1} \\
\delta^{(1)} A_{\mu} &=& i\bar{\epsilon}
     \Gamma^{\mu}\psi,
\end{Eqnarray}
and
\begin{Eqnarray}
\delta^{(2)} \psi &=& \xi  \label{susy2}\\
\delta^{(2)} A_{\mu} &=& 0,
\end{Eqnarray}
and translation symmetry:
\begin{equation}
\delta A_{\mu} = c_{\mu} {\bf 1}.
\end{equation}
If we interpret the eigenvalues of $A_{\mu}$ 
as the space-time coordinates, 
we can regard the above symmetry as ${\cal N}=2$ supersymmetry. 
We take a linear combination of $\delta^{(1)}$ and $\delta^{(2)}$ as 
\begin{Eqnarray}
\tilde{\delta}^{(1)} &=& \delta^{(1)} + \delta^{(2)}   \cr
\tilde{\delta}^{(2)} &=& 
   i\left(\delta^{(1)} - \delta^{(2)} \right) ,
\end{Eqnarray}
we can obtain ${\cal N}=2$ supersymmetry algebra:
\begin{Eqnarray}
\left(
\tilde{\delta}_{\epsilon}^{(i)}\tilde{\delta}_{\xi}^{(j)}
 -\tilde{\delta}_{\xi}^{(j)}\tilde{\delta}_{\epsilon}^{(i)}
\right) \psi &=& 0, \cr
\left(
\tilde{\delta}_{\epsilon}^{(i)}\tilde{\delta}_{\xi}^{(j)}
 -\tilde{\delta}_{\xi}^{(j)}\tilde{\delta}_{\epsilon}^{(i)}
\right) A_{\mu} &=& 2i \bar{\epsilon} \Gamma^{\mu} \xi \delta_{ij}
\end{Eqnarray}
The equation of motion of (\ref{action}) 
with $\psi=0$ is 
\begin{equation} 
 \left[ A_{\mu} , \left[ A_{\mu} ,A_{\nu}\right] \right] =0. 
\label{eom} 
\end{equation}
The solution in which 
$\left[ A_{\mu} ,A_{\nu}\right]= c\mbox{-number}$ 
is an interesting solution 
because it is a BPS background. 
That is, half of the supersymmetry 
is preserved in this background. 
This corresponds to a D-brane background. 

\vspace{0.4cm}
We expand the theory around the following classical solution, 
\begin{equation}
\left[ \hat{p}^{\mu} , \hat{p}^{\nu} \right] = i B^{\mu \nu},  
\end{equation} 
where $B_{\mu \nu}$ is  anti-symmetric tensor and $c$-number. 
This is a solution of (\ref{eom}) and 
is a BPS background.  
We assume the rank of $B_{\mu \nu}$ to be $\tilde{d}$ 
and define its inverse $C^{\mu \nu}$ 
in $\tilde{d}$ dimensional subspace. 
$\hat{p}^{\mu}$ satisfy the canonical commutation relations 
and span the $\tilde{d}$ dimensional phase space. 
The volume of the phase space is 
$V_{p} = n(2\pi)^{\tilde{d}/2} \sqrt{detB}$. 

We expand $A_{\mu}=\hat{p}_{\mu}+\hat{a}_{\mu}$ and 
Fourier-decompose $\hat{a}_{\mu}$ and $\hat{\psi}$ as 

\begin{equation}
\hat{a}_{\mu}=\sum _{k} \tilde{a}_{\mu}(k) 
\exp(i C^{\mu \nu} k_{\mu} \hat{p}^{\nu}),
\end{equation}

\begin{equation}
\hat{\psi}=\sum _{k} \tilde{\psi}(k) 
\exp(i C^{\mu \nu} k_{\mu} \hat{p}^{\nu}).
\end{equation}
$\exp(i C^{\mu \nu} k_{\mu} \hat{p}^{\nu})$ is the eigenstate 
of $P_{\mu}=[\hat{p}^{\mu}, \cdot  ]$  with eigenvalue $k_{\mu}$.
The Hermiticity requires that 
$\tilde{a}_{\mu}^{\ast}(k) = \tilde{a}_{\mu}(-k)$
and $\tilde{\psi}_{\mu}^{\ast}(k) =\tilde{\psi}_{\mu}(-k)$. 
Let $\Lambda$ be the extension of each $\hat{p}_{\mu}$.
The volume of one quantum in this phase space is 
$\Lambda^{\tilde{d}} /n=\lambda^{\tilde{d}}$
where $\lambda$ is the spacing of the quanta, 
say, noncommutative scale. 
$B$ , which is the component of $B_{\mu\nu}$, 
is related to $\lambda$ as $B=\lambda^{2}/2\pi$.
$k^{\mu}$ is quantized 
in the unit of $k_{\mu}^{min}=\Lambda/n^{2/\tilde{d}}=\lambda/n^{1/\tilde{d}}$.
The range of $k_{\mu}$ is restricted as 
$-n^{1/\tilde{d}} \lambda/2 \le k_{\mu} 
\le n^{1/\tilde{d}} \lambda/2.$
% So $\sum _{k}$ runs over $n^2$ degrees of freedom 
% which coincide with those of $n$ dimensional hermitian matrices.

Consider the map from a matrix to a function as 

\begin{equation}
\hat{a}_{\mu} \rightarrow 
a_{\mu}(x)=\sum _{k} \tilde{a}_{\mu}(k) 
\exp(i k_{\mu} x^{\mu}).  \label{map}
\end{equation}
Under this map, we obtain the following map, 
\begin{equation}
\hat{a}\hat{b} \rightarrow 
a(x) \star b(x), 
\end{equation}
where $\star$ is the star product defined as follows, 

\begin{equation} 
a(x) \star b(x) \equiv 
\exp \left( \frac{i C^{\mu\nu}}{2} 
\frac{\partial^2}{\partial \xi^{\mu} \partial \eta^{\nu}} \right)
a(x+\xi)b(x+\eta) \mid _{\xi=\eta=0}.    \label{star}
\end{equation}
$Tr$ over matrices can be mapped 
on the integration over functions as 

\begin{equation}
Tr[\hat{a}] =\sqrt{detB} 
\left(\frac{1}{2\pi} \right)^{\frac{\tilde{d}}{2}}
\int d^{\tilde{d}}x a(x) .
\end{equation}
Using these rules, 
the adjoint operator of $\hat{p}_{\mu}+\hat{a}_{\mu}$ is mapped 
to the covariant derivative:
\begin{equation}
\left[\hat{p}_{\mu} + \hat{a}_{\mu},\hat{o} \right] 
\rightarrow \frac{1}{i} \partial_{\mu} o(x) 
+a_{\mu}(x) \star o(x) - o(x) \star a_{\mu}(x)
\equiv \left[D_{\mu},o(x) \right].
\end{equation}

Noncommutative Yang-Mills action can be obtained 
by applying the above rules. 
Although we have discussed the momentum space, 
the coordinate space is also embedded in the matrices 
of twisted reduced model through the relation 
$\hat{x}^{\mu}=C^{\mu\nu}\hat{p}_{\nu}$.
This relation says that 
the coordinate space is related to the momentum space.
This relation reminds us T-duality.

%%%%%%%%%%%%%%%%%%%%%%%%%%%%%%

\section{The Gauge theory of D1-D5 system in IIB matrix model}

\hspace{0.4cm}
Following the rules of the previous section, 
we consider IIB matrix model with D1-D5 backgrounds.
We use the coordinate space description 
instead of the dual momentum space.
D1-D5-brane solutions are constructed 
by preparing the following matrices\cite{IKKT,CMZ,FS}, 

\begin{Eqnarray}
 \hat{x}^{0} &=&\frac{T}{\sqrt{2\pi n_{1}}} 
          \hat{q}  \cr
 \hat{x}^{1} &=&\frac{L_1}{\sqrt{2\pi n_{1}}} 
          \hat{p}  \cr
 \hat{x}^{2} &=&\frac{L_2}{\sqrt{2\pi n_{2}}} 
          \hat{q}^{\prime}  \cr
 \hat{x}^{3} &=&\frac{L_3}{\sqrt{2\pi n_{2}}} 
          \hat{p}^{\prime}  \cr
 \hat{x}^{4} &=&\frac{L_4}{\sqrt{2\pi n_{3}}} 
          \hat{p}^{\prime \prime}  \cr
 \hat{x}^{5} &=&\frac{L_5}{\sqrt{2\pi n_{3}}} 
          \hat{p}^{\prime \prime}  .
\end{Eqnarray}
where $\left[ \hat{q} , \hat{p} \right] = i$, 
$\left[ \hat{q}^{\prime} , \hat{p}^{\prime} \right] = i$, 
$\left[ \hat{q}^{\prime \prime} , 
\hat{p}^{\prime \prime} \right] = i$.
D1-brane and D5-brane are constituted of 
$n_{D1} (=n_{1})$ quanta of the volume $l_{NC}^{2}$ 
and $n_{D5} (=n_{1}n_{2}n_{3})$ quanta of the volume $l_{NC}^{6}$ 
respectively, 
where $TL_{1}=n_{D1} l_{NC}^{2}$ and 
$TL_{1}L_{2}L_{3}L_{4}L_{5}=n_{D5} l_{NC}^{6}$.  
$l_{NC}=2\pi/\lambda$ is the noncommutative length scale.

$\hat{x}^{\mu}$ satisfy the commutation relation such as 
\begin{equation}
\left[ \hat{x}^{\mu} , \hat{x}^{\nu} \right] 
   = -i C^{\mu \nu}, \label{comm} 
\end{equation}
where $C^{01}=C^{23}=C^{45}=2\pi/B=l_{NC}^{2}
 =(2\pi/\lambda)^{2}$.
In IIB matrix model, 
the D-brane is interpreted as not pure one but 
one with 
non-vanishing gauge field strength $B_{\mu\nu}$\cite{CT}.
%(Therefore, it contains the lower-dimensional D-branes. )
They are equivalent to the branes 
in the presence of a constant 
NS-NS two-form background $b_{\mu\nu}$
\footnote{This is the configuration considered in \cite{SW}}.

We expand the model around 
the classical background, $A_{\mu}^{cl}$, 
and the fluctuation is denoted by $X_{\mu}$,
\begin{equation}
A_{\mu}=A_{\mu}^{cl}+X_{\mu}.
\end{equation}
We align a D1-brane along the $0,1$ directions and 
a D5-brane along the $0,1,2,3,4,5$ directions:
\begin{Eqnarray}
A_{\mu}^{cl}  
 &=&   \left( \begin{array}{c c}
  \hat{x}_{\mu} & 0 \\ 0  &  \hat{x}_{\mu}   \\
 \end{array} \right)     \hspace{0.4cm}  (\mu =0,1) \cr
&=& \left( \begin{array}{c c}
  0 & 0 \\ 0  &  \hat{x}_{\mu}   \\
 \end{array} \right)  \hspace{0.4cm}  (\mu=2,3,4,5) \cr
&=& \left( \begin{array}{c c}
  u_{\mu} & 0 \\ 0  &  0  \\
 \end{array} \right)  \hspace{0.4cm}  (\mu=6,7,8,9)   
           \label{background}  \\
X_{\mu}  
 &=&   \left( \begin{array}{c c}
  \hat{a}_{\mu} & \hat{c}_{\mu} 
  \\ \hat{c}^{\dagger}_{\mu}  &  \hat{b}_{\mu}  \\
 \end{array} \right).  
\end{Eqnarray}
These $A_{\mu}^{cl}$ satisfy the equation of motion (\ref{eom}).
The upper left part and the lower right part of matrices 
represent a D1-brane and a D5-brane respectively.
The remaining parts of matrices represent 
the interactions of them.
$u_{\mu}$ are the positions of D1-branes in the 6,7,8,9 directions.
We will use the indices $\alpha$, $\beta$ 
as the indices running over the direction parallel 
to the branes 
and $a$, $b$ as the indices over the direction 
transverse to the branes.
We replace $\hat{a}_{\alpha}$ and $\hat{b}_{\alpha}$ with 
$C^{\alpha \beta}a_{\beta}(x)$ and $C^{\alpha \beta}b_{\beta}(x)$ 
in our correspondence (\ref{map}), 
where $a_{\alpha}(x)$ and $b_{\alpha}(x)$ can be interpreted 
as gauge fields.
We also replace $\hat{a}_{a}$, $\hat{b}_{b}$ and $\hat{c}_{\mu}$ with 
$(1/\sqrt{B})a_{a}(x)$, $(1/\sqrt{B})b_{b}(x)$ and $(1/\sqrt{B})c_{\mu}(x)$.

We can consider multiple branes by replacing 
$\hat{x} \rightarrow  \hat{x} \otimes {\bf 1} _{Q_{1}}$ 
or $\hat{x} \otimes {\bf 1} _{Q_{5}}$.
$Q_1$ and $Q_5$ are the number of D1-branes and D5-branes respectively.
We call this system D1-D5 system.
This system is described by 
two-dimensional\footnote{ 
To describe the compact space, 
we impose the following condition for the 
matrices, 
\begin{displaymath}
A_{\mu}+ 2\pi R_{\mu} \delta_{\mu \nu}
 = \Omega_{\nu} A_{\mu} \Omega_{\nu}^{\dagger}
\hspace{0.7cm}(\mu,\nu=2,3,4,5) 
\end{displaymath}
where $\Omega_{\mu}$ are unitary matrices 
and $R_{\mu}$ are compactification radii. 
$\Omega_{\mu}$ don't commute for the noncommutative torus
\cite{CDS}.} 
supersymmetric 
$U(Q_{1}) \times U(Q_{5})$ gauge theory in low energy limit. 
$a_{\mu}$ are $Q_{1} \times Q_{1}$ hermitian matrices 
and transform as adjoints of $U(Q_{1})$.
$b_{\mu}$ are $Q_{5} \times Q_{5}$ hermitian matrices 
and transform as adjoints of $U(Q_{5})$.
$c_{\mu}$ are $Q_{1} \times Q_{5}$ matrices 
and transform as 
the product of the fundamental representation of $U(Q_{1})$
and the anti-fundamental representation of $U(Q_{5})$.
$c_{\mu}^{\dagger}$ are the complex conjugate of $c_{\mu}$.
$a_{\mu}$, $c_{\mu}$ and $c_{\mu}^{\dagger}$ are 
two-dimensional fields while $b_{\mu}$ are 
six-dimensional fields. 

We examine the condition whether this background is supersymmetric:
\begin{Eqnarray}
\delta^{(1)} \psi &=& \frac{i}{2} 
  \left[ A_{\mu} ,A_{\nu}\right] \Gamma^{\mu\nu} \epsilon \cr
 &=& 
  \left( \begin{array}{c c}
  C^{01}\Gamma^{01} & 0 \\ 0  &  
    C^{01} \Gamma^{01} 
   +C^{23}\Gamma^{23}+C^{45}\Gamma^{45}   \\
 \end{array} \right)  \epsilon ,  \cr
\delta^{(2)} \psi &=& \xi.  \label{susytr}
\end{Eqnarray}
We find that the following conditions are needed 
for this configuration to be supersymmetric: 

\begin{displaymath}
C^{01}\Gamma^{01}\epsilon =\xi, 
\end{displaymath}
\begin{equation}
 \Gamma^{2}\Gamma^{3}\Gamma^{4}\Gamma^{5} \epsilon 
  =\epsilon . \label{susy4}
\end{equation}
Therefore one fourth of  
the supersymmetry are preserved in this configuration. 
This theory has ${\cal N}=(4,4)$ supersymmetry 
in two dimensions. 

We now consider twisted reduced model action with 
the D1-D5 backgrounds. 
The bosonic part action with $\hat{c}_{\mu}=0$ 
can be easily calculated and become 

\begin{Eqnarray}
S &=& \frac{TLQ_{1}}{4\pi g^{2}B^{2}} 
  - \frac{1}{2\pi B^{3}g^2} \int d^{2}x tr 
   ( [ D_{\alpha},D_{\beta}] 
      [ D_{\alpha}, D_{\beta}]    \cr
 &+& 2B\left( D_{\alpha} a _{a} \right)\left( D_{\alpha} a _{a} \right) 
   + B^{2} [a_{a},a_{b}] [a_{a},a_{b}]    
    ) _{\star}  \cr 
   &+& \frac{3BTL^{5}Q_{5}}{16\pi^{3} g^{2}} 
  - \frac{1}{2\pi^{3} g^{2} B} \int d^{6}x tr  
  ( [ \tilde{D}_{\alpha},\tilde{D}_{\beta}] 
      [ \tilde{D}_{\alpha},\tilde{D}_{\beta}]   \cr 
   &+& 2B \left( \tilde{D}_{\alpha} b_{a} \right)
      \left( \tilde{D}_{\alpha} b_{a} \right)     
   +B^{2} [b_{a},b_{b}][b_{a},b_{b}] 
        )_{\star},
\end{Eqnarray}
where $D_{\alpha}$ is the covariant derivative 
constructed by the gauge fields $a_{\alpha}(\alpha=0,1)$ 
while $\tilde{D}_{\alpha}$ is the covariant derivative 
constructed by the gauge fields 
$b_{\alpha}(\alpha=0,1,2,3,4,5)$.
$tr$ is taken over the color indices.
$\star$ means that products are not usual ones 
but the ones defined by (\ref{star}).
We have just considered the situation that D1-branes are 
overlapping and also that D5-branes are overlapping. 
By replacing left upper part of $A_{2}^{cl}$ with 
$w_{2}^{(i)} \otimes {\bf 1} _{Q_{1}}$ ($i =$ color index) 
which is the positions of D1-branes 
in the $x_{2}$ direction, 
we can obtain the case that   
D1-branes are separated in $x_{2}$ direction. 
This makes $a_{\mu}$ field massive. 
Other cases can be done in the same way.

We next consider 
$c_{\mu}$ and $c_{\mu}^{\dagger}$ fields. 
It corresponds to the states appearing in 
$(1,5)$ and $(5,1)$ 
string\footnote{$(1,5)$ string 
is the open string with one end on a D1-brane and 
the other end on a D5-brane.} 
in string theory picture. 
We ignore the interaction terms.
After the gauge fixing\cite{IKKT}, 
the kinetic and mass terms are given by 

\begin{Eqnarray}
S =&-& 
\frac{1}{8\pi B^2 g^2} \int d^{2}x tr
 \sum _{\alpha=0}^{1} \sum_{\mu=0}^{9}
 \{ \partial_{\alpha} c_{\mu} \partial_{\alpha} c_{\mu}^{\dagger}  
   +\partial_{\alpha} c_{\mu}^{\dagger} \partial_{\alpha} c_{\mu}
 \} _{\star} \cr
&-& 
\frac{1}{8\pi g^2} \int d^{2}x tr
 \sum _{\alpha=2}^{5} \sum_{\mu=0}^{9}
 \{ c_{\mu} (\hat{x}_{\alpha}^{2} c_{\mu}^{\dagger})  
   + U^{2} c_{\mu}c_{\mu}^{\dagger}
    +c_{\mu}^{\dagger} (c_{\mu}\hat{x}_{\alpha}^{2} ) 
     +U^{2} c_{\mu}^{\dagger} c_{\mu}
\} _{\star} \cr
&+& i\frac{1}{8\pi Bg^2}  \int d^{2}x tr 
 \{c_{2}c_{3}^{\dagger} -c_{2}^{\dagger}c_{3}
  -c_{3}c_{2}^{\dagger} +c_{3}^{\dagger}c_{2}
  +c_{4}c_{5}^{\dagger} -c_{4}^{\dagger}c_{5}
  -c_{5}c_{4}^{\dagger} +c_{5}^{\dagger}c_{4} 
  \} _{\star} , \label{caction}
\end{Eqnarray}
where $U^{2} \equiv \sum_{\mu =6}^{9} u_{\mu}^{2}$.
%where $tr U^{2} c_{\mu}c_{\mu}^{\dagger}
% \equiv ( u_{\mu})^{2}
% (c_{\mu})_{ij}(c_{\mu}^{\dagger})_{ji}$ 
%($i,j$ = color indices).  
Mass terms are given in second and third lines in (\ref{caction}). 
$\hat{x}_{2}^{2}+\hat{x}_{3}^{2}$ 
is the Hamiltonian of the charged particle 
moving on the two-plane in a uniform magnetic field 
(remember (\ref{comm})) .
Therefore the energy eigenvalue of 
$\hat{x}_{2}^{2}+\hat{x}_{3}^{2}$ 
is given by the Landau level,
$2(N+ \frac{1}{2})/B$ ( $N=0,1,2,\ldots$) 
and similarly for $\hat{x}_{4}^{2}+\hat{x}_{5}^{2}$.
Now we can rewrite mass terms in (\ref{caction}) as 

\begin{displaymath}
-\frac{1}{8\pi g^{2}} \int d^{2}x tr 
\left[ ((2M+2)/B+U^{2}) \left(\phi_{1} \phi_{1}^{\dagger}
  +\phi_{1}^{\dagger} \phi_{1} \right)
 +(2M/B+U^{2})\left(\phi_{2} \phi_{2}^{\dagger}
  +\phi_{2}^{\dagger} \phi_{2} \right) 
     \right]_{\star} 
\end{displaymath}
\begin{displaymath}
+\frac{1}{8\pi g^{2} B^{2}} \int d^{2}x tr 
\left[ ((2N+2)/B+U^2) \left(\phi_{3} \phi_{3}^{\dagger}
  +\phi_{3}^{\dagger} \phi_{3} \right)
 +(2N/B+U^2)\left(\phi_{4} \phi_{4}^{\dagger}
  +\phi_{4}^{\dagger} \phi_{4} \right) 
     \right]_{\star}   \label{cmass}
\end{displaymath}
\begin{equation}
-\frac{1}{8\pi g^{2} } \int d^{2}x tr 
  \sum_{\mu \neq 2,3,4,5}
\left[ ((2M+1)/B+(2N+1)/B+U^{2}) 
      \left(c_{\mu} c_{\mu}^{\dagger} 
    +c_{\mu}^{\dagger} c_{\mu} \right)
  \right]_{\star} .
\end{equation}
where $\phi_{1} =\frac{1}{\sqrt{2}}(c_{2}+ic_{3})$, 
  $\phi_{2} =\frac{1}{\sqrt{2}}(c_{2}-ic_{3})$, 
$\phi_{3} =\frac{1}{\sqrt{2}}(c_{4}+ic_{5})$ 
and $\phi_{4} =\frac{1}{\sqrt{2}}(c_{4}-ic_{5})$. 
$U^2$ is the distance between D1-branes and D5-branes.
These mass terms mean that a field 
appearing in the string connecting two branes has a mass 
proportional to the distance between two branes 
in string picture. 
We can say 
that only $\phi_{2}$ and $\phi_{4}$ 
can be massless scalar 
at the ground state ($M=0$ or $N=0$) and 
$U^2=0$ 
%(this means that D1-branes and D5-branes are overlapping) 
while $\phi_{1}$,$\phi_{3}$ and $c_{\mu}(\mu \neq 2,3,4,5)$ 
are always massive 
because of the zero point energy. 
Only four real degrees of freedom are 
massless degrees of freedom. 

\vspace{0.4cm}
We next examine the fermionic part. 
We divide matrices into four parts in the same way 
as the bosonic part.
Fermionic background is set to zero.
\begin{equation}
\hat{\psi}  
 =   \left( \begin{array}{c c}
  \hat{\psi}_{1} &  \hat{\psi}_{3}  \\  
  \hat{\psi}_{3}^{\dagger} &  \hat{\psi}_{2}   \\
 \end{array} \right)      \label{fermion}
\end{equation}
Kinetic and mass terms are calculated as follows, 
\begin{Eqnarray}
S= &-&\frac{1}{4\pi g^2} \int d^{2}x tr \left(
\bar{\psi}_{1} \tilde{\Gamma}_{\alpha}
                      \partial_{\alpha} \psi_{1}       
        \right)_{\star}  \cr
&-&   \frac{1}{4\pi g^2} \int d^{2}x tr (
\bar{\psi}_{3} \tilde{\Gamma}_{\alpha} 
              \partial_{\alpha} \psi^{\dagger}_{3} 
+ \bar{\psi_{3}^{\dagger}} \tilde{\Gamma}_{\alpha} 
        \partial_{\alpha} \psi_{3}   \cr
&&+ B \bar{\psi}_{3} \left(   
   \Gamma_{\gamma} \hat{x}^{\gamma} 
   -u_{\tau} \Gamma_{\tau}  \right) 
\psi^{\dagger}_{3} 
+ B \bar{\psi_{3}^{\dagger}} \left(
   \Gamma_{\gamma} \hat{x}^{\gamma} 
    + u_{\tau} \Gamma_{\tau}
       \right)\psi_{3}    
         )_{\star}      \cr
&-& \frac{B^2}{16\pi^3 g^2} \int d^{6}x tr \left(
      \bar{\psi}_{2} \tilde{\Gamma}_{\beta} 
        \partial_{\beta} \psi_{2} 
      \right)_{\star},   \label{fermionmass}
\end{Eqnarray}
where $\tilde{\Gamma}_{\beta}
   =\Gamma_{\delta} \epsilon^{\delta \beta}$,  
%and $trU^{1}\bar{\psi}_{1} \psi_{1}
%=(u^{(k)}-u^{(i)})(\bar{\psi}_{1})_{ik}(\psi_{1})_{ki}$, 
%$trU^{2}\bar{\psi}_{2} \psi_{2}
%=(v^{(k)}-v^{(i)})(\bar{\psi}_{2})_{ik}(\psi_{2})_{ki}$, 
%$trU^{3}\bar{\psi}_{3} \psi_{3}^{\dagger}
%=(u^{(k)}-v^{(i)})(\bar{\psi}_{3})_{ik}(\psi_{3}^{\dagger})_{ki}$, 
%and 
%$trU^{3}\bar{\psi}_{1} \psi_{1}
%=(v^{(k)}-u^{(i)})(\bar{\psi^{\dagger}}_{3})_{ik}(\psi_{3})_{ki}$. 
and $\alpha=0,1$, $\beta=0,1,2,3,4,5$, $\gamma=2,3,4,5$, 
and $\tau= 6,7,8,9$. 
%We notice that $\psi_{1}$ and $\psi_{2}$ 
%are always massless.
We consider the massless conditions 
         for $\psi_{3}$ and $\psi_{3}^{\dagger}$. 
Mass terms are given by the fourth and fifth terms 
of (\ref{fermionmass}).  
Therefore 
the massless conditions 
for $\psi_{3}$ and $\psi_{3}^{\dagger}$ 
are  
$\Gamma_{\gamma} \hat{x}^{\gamma} \psi_{3} =0$ and 
$\Gamma_{\gamma} \hat{x}^{\gamma} \psi_{3}^{\dagger} =0$ 
and $u_{\tau}=0$.
We rewrite $\Gamma_{\gamma} \hat{x}^{\gamma} =0$ as
\begin{Eqnarray}
\left( \Gamma_{2}+i\Gamma_{3} \right)
       \left(\hat{x}^{2}-i\hat{x}^{3} \right)
&+&\left( \Gamma_{2}-i\Gamma_{3} \right)
 \left(\hat{x}^{2}+i\hat{x}^{3} \right)  \cr
+&&\left( \Gamma_{4}+i\Gamma_{5} \right)
       \left(\hat{x}^{4}-i\hat{x}^{5} \right)
+\left(\Gamma_{4}-i\Gamma_{5} \right)
 \left(\hat{x}^{4}+i\hat{x}^{5} \right)  =0.
\end{Eqnarray}
From the commutation relation (\ref{comm}), 
we can make annihilation-creation operators:
\begin{Eqnarray}
a_{23} & \equiv & \sqrt{\frac{B}{2}} 
       \left(\hat{x}^{3}+i\hat{x}^{2}\right) \cr
a_{23}^{\dagger} & \equiv & \sqrt{\frac{B}{2}} 
       \left(\hat{x}^{3}-i\hat{x}^{2}\right)
\end{Eqnarray}
\begin{equation}
\left[a_{23},a_{23}^{\dagger}\right]=1.
\end{equation}
Annihilation-creation operators for 
$\hat{x}^{4}$ and $\hat{x}^{5}$ also can be made 
in the same way.
So the following conditions are needed for 
$\psi_{3}$ and $\psi_{3}^{\dagger}$ to be massless:
\begin{Eqnarray}
a_{23}  | \psi_{30} \rangle  &=& 0   \cr
a_{45}  | \psi_{30} \rangle  &=& 0   \cr
(\Gamma_{2}+i\Gamma_{3}) |a_{23}^{\dagger} 
           \psi_{30} \rangle  &=&0  \cr
(\Gamma_{4}+i\Gamma_{5}) |a_{45}^{\dagger} 
     \psi_{30} \rangle  &=&0  \label{gamma45}.   
\end{Eqnarray}
Last two conditions say that 
one fourth of $\psi_{3}$ components survive.
In other words, 
the massless mode of $\psi_{3}$ fermion 
has eight components. 
We have seen that massless mode of $c_{\mu}$ 
has real four degrees of freedom.
Both $c_{\mu}$ and $\psi_{3}$ transform 
as the product of the fundamental of $U(Q_{1})$ 
and anti-fundamental of $U(Q_{5})$, 
i.e. these are super-partners.
Since the massless 
physical degrees of freedom of $\psi_{3}$ 
are four, they are same with 
the massless degrees of freedom of $c_{\mu}$. 

\vspace{0.4cm}

We compare these results with string theory results. 
Off-diagonal elements of matrices 
correspond to $(1,5)$ or $(5,1)$ string. 
Superstring theory is usually formulated in the 
Ramond-Neveu-Schwarz formalism, while 
matrix model is formulated in the Green-Schwarz formalism. 
We examine the spectrum of $(1,5)$-string 
in the presence of $b_{\mu\nu}$ field 
by the Ramond-Neveu-Schwarz formalism. 
In the R sector, the zero point energy is zero, 
because world-sheet bosons and world-sheet fermions 
have the same moding. 
The massless spectrum is not affected 
by a $b_{\mu\nu}$ field.
There are four periodic world-sheet fermions 
in the NN and DD directions 
(0,1 directions are not counted ). 
So, there are four zero modes.  
These zero modes generate $2^{4/2}=4$ ground states. 
After imposing the GSO projection, 
two of the four ground states survive. 
There are also the contribution from $(5,1)$ string. 
Therefore there appear four massless modes, 
which correspond to the massless modes of 
$\phi_{2}$ and $\phi_{4}$, 
from the R sector. 
In the NS sector, 
the massless spectrum is affected  
by the $b_{\mu\nu}$ field. 
The zero point energy has the dependence 
of the value of the $b_{\mu\nu}$ field. 
The spectrum is discussed in \cite{SW} about 
the case of the D0-D4 system. 
Same discussions are applied to the D1-D5 system.
For large $b_{23}$ and $b_{45}$, a D5-brane is regarded 
as to be constituted of infinitely D1-branes or anti-D-branes.
According to \cite{SW}, 
the spectrums are different in these cases. 
(From the open string stretching between D-string and 
anti-D-string, tachyon state appears. 
So this system is unstable and not supersymmetric.) 
The case which is considered in this paper 
is the former case in which this system is supersymmetric. 
In this case, total four massless modes, 
which correspond to the massless modes of 
$\psi_{3}$, 
appear from the NS sector 
after imposing the GSO projection. 
From these facts, 
we conclude that 
the number of massless modes for the off-diagonal elements 
in matrix model matches the number of massless modes 
for (1,5) and (5,1) string.

\vspace{0.8cm}

We now consider the description of D1-branes 
embedded in D5-branes. 
We are interested in D1-D5 bound states  
because these states are associated with 
the dynamics of the black hole\cite{SV}. 
It can be shown that the instantons 
of $U(Q_{5})$ gauge theory on the D5-branes
carry R-R two-form charge and 
D1-branes considered in the previous part of this section 
correspond to the zero size limit of the instantons.
Therefore $Q_{1}$ D1-branes embedded in D5-brane are 
constructed as $Q_{1}$ instantons 
in $U(Q_{5})$ gauge theory\cite{Doug,costa}. 
We now look at the Chern-Simon coupling in D5-brane action: 
\begin{displaymath} 
tr \int _{\Sigma_{6}} C_{2}\wedge F_{2} \wedge F_{2} 
\end{displaymath} 
where $C_{2}$ is the R-R two form field and 
$F_{2}$ is the gauge field strength. 
In this coupling, $F\wedge F$ plays the role of D1-brane charge.
In general, lower-dimensional D-branes embedded in 
higher-dimensional D-branes 
are expressed as the magnetic flux 
on higher-dimensional D-branes\cite{Taylor}. 

We consider $U(Q_5)$ gauge theory on D5-branes
with a self-dual gauge field strength on $T^4$ in matrix model. 

\begin{Eqnarray}
A_{\mu}^{cl}  
 &=&   \left( \begin{array}{c c}
  \hat{x}_{\mu} & 0 \\ 0  &  \hat{x}_{\mu}   \\
 \end{array} \right)     \hspace{0.4cm}  (\mu =0,1,2,3,4,5) \cr
&=& \hspace{0.4cm} 0 \hspace{0.8cm}  (\mu=6,7,8,9) \\
X_{\mu}  
 &=&   \left( \begin{array}{c c}
  \hat{b}_{\mu}^{(1)} & \hat{b}_{\mu}^{(3)} 
  \\ \hat{b}_{\mu}^{(3)\dagger}  &  \hat{b}_{\mu}^{(2)}  \\
\end{array} \right).  
\end{Eqnarray}
%\begin{equation}
%A_{\mu} =\hat{x}_{\mu}+\hat{b}_{\mu} 
%     \hspace{0.4cm}(\mu= 0,1,2,3,4,5) 
%\end{equation} 
The size of the right lower part of matrices are $2 \times 2$.
The size of the left upper part of matrices 
are $(Q_{5}-2) \times (Q_{5}-2)$.
The equation of motion in IIB matrix model 
is given by (\ref{eom}).
After the replacements from matrices to functions, 
the equation of motion of this gauge theory become 

\begin{equation}
 \left[ \tilde{D}_{\mu} , 
 \left[ \tilde{D}_{\mu} , 
    \tilde{D}_{\nu}\right] \right] =0 \hspace{0.5cm} 
         (\mu,\nu=0,1,2,3,4,5,6). 
\end{equation}
We analyze the situation that only gauge fields $b_{\mu}^{(2)}$
have nontrivial classical solutions.
In other words, we embed an (anti-)instanton into 
$SU(2)$ part (right lower part). 
The instanton solution
\footnote{
Instantons on non-commutative ${\bf R^{4}}$ space 
are constructed in \cite{NS}.} 
can be constructed as follows, 
\begin{Eqnarray}
\tilde{F}^{(2)}_{\alpha \beta} 
      &=& \pm \ast \tilde{F}^{(2)}_{\alpha \beta} \cr
\tilde{F}^{(2)}_{\rho \sigma} 
       &=& 0 \cr
\tilde{F}^{(2)}_{\rho \alpha} &=& 0 ,       \label{ins}
\end{Eqnarray}
where $\alpha, \beta= 2,3,4,5$ and $\rho, \sigma= 0,1$. 
$\ast$ means the dual tensor on $T^{4}$.
The star product which is defined in (\ref{star}) 
is used in these equations. 
$\tilde{F}^{(2)}$ is the gauge field strength constructed 
by the gauge field $b^{(2)}$.
Self-dual and anti-self-dual 
configurations are referred to instantons 
and anti-instantons respectively. 
In D-brane picture, the winding number of instanton 
represents the D1-brane charge.
Therefore to express $Q_{1}$ D1-branes on D5-branes, 
we set the following condition: 
\begin{equation} 
Q_{1}=\frac{1}{8\pi^2} tr \int_{T^4} 
\tilde{F}^{(2)} \wedge \tilde{F}^{(2)}. 
    \label{wind}
\end{equation} 
%where $n$ is the winding number of instanton. 
Also, D5-branes carry D3-branes charge. 
Therefore we also need the following condition 
to obtain pure D1-D5 bound states, i.e. no D3-branes: 
\begin{equation}
\frac{1}{2\pi} tr \int_{T^4} \tilde{F}^{(2)}=0. 
\end{equation}
In these configurations, 
supersymmetry transformations (\ref{susy1}) and (\ref{susy2}) 
are given as follows:
\begin{Eqnarray}
\delta^{(1)} \psi &=& \frac{i}{2} 
  \left[ A_{\mu} ,A_{\nu}\right] \Gamma^{\mu\nu} \epsilon \cr 
&=&  \frac{1}{2} \left( \begin{array}{c c}
  C^{\mu\nu}\Gamma^{\mu\nu} & 0 \\ 0  & 
     \left( C^{\mu\nu} +iC^{\mu \tau}C^{\nu \rho} 
   \left[\tilde{D}^{(2)}_{\tau},\tilde{D}^{(2)}_{\rho} \right] \right)  
   \Gamma^{\mu\nu}   \\
 \end{array} \right) \epsilon    \cr
&=& \frac{1}{2}C^{\mu\nu}\Gamma^{\mu\nu}\epsilon  \cr
&&+\left( \begin{array}{c c}
  0 & 0 \\ 0  & 
      iC^{\alpha \gamma}C^{\beta \delta} 
   \left[D_{\gamma},D_{\delta} \right] 
 \left( \frac{1 \mp \Gamma^{2}\Gamma^{3}\Gamma^{4}\Gamma^{5}}{2} \right) \\
\end{array} \right) \Gamma^{\alpha \beta} \epsilon \cr
\delta^{(2)} \psi &=& \xi,
\end{Eqnarray}
where repeated indices $\alpha$,$\beta$,$\gamma$,$\tau$ 
run over 2,3,4,5.
$\mp$ correspond to self-dual and anti-self-dual configurations 
respectively. 
Following constraints are needed for this configuration 
to be supersymmetric: 
\begin{displaymath}
  \frac{1}{2}C^{\mu\nu}\Gamma^{\mu\nu}\epsilon
=\xi,
\end{displaymath}
\begin{equation}
 \Gamma^{2}\Gamma^{3}\Gamma^{4}\Gamma^{5} \epsilon
 =\pm \epsilon .        \label{susy5}
\end{equation}

We study the zero mode.
It is known that 
the second Chern number is equal to the difference in 
the number of the chiral fermion zero modes 
whose chiralities are positive and negative (the index theorem).
The difference of the chiral fermions is given 
by $\alpha_{V}Q_{1}$\cite{schwarz}, 
where $\alpha_{V}$ is the Dynkin index 
depending on the representation and $Q_{1}$ is defined in (\ref{wind}). 
We expect that chiral fermions appear 
from the off-diagonal parts and the right lower part. 
These parts transform as the fundamental and 
adjoints of $SU(2)$ respectively. 
In these cases, $\alpha_{V}$ is 1 and 4 respectively.
Therefore $Q_{1}Q_{5}$ chiral fermion zero modes 
appear from the index theorem.
From the condition (\ref{gamma45}), 
the zero mode of $\psi_{3}$ satisfies 
$\Gamma_{2}\Gamma_{3}\Gamma_{4}\Gamma_{5}\psi_{30}=-\psi_{30}$. 
This implies that $\psi_{30}$ has negative chirality as 
a $SO(4)$ (the rotation of $x_{2}$, $x_{3}$, $x_{4}$, $x_{5}$)
Weyl spinor. 
We notice that this chirality is opposite to 
the spinor which satisfy 
the condition of unbroken supersymmetry (see (\ref{susy5})). 
The reason is that these zero modes are Goldstone modes. 
The degrees of freedom of these zero modes 
is associated with the dynamics of the black hole entropy\cite{SV}. 

%%%%%%%%%%%%%%%%%%%%%%%%%%%%%%

\section{Relation to supergravity solution} 
\hspace{0.4cm} 
In this section, we would like 
to study the supergravity solution  
in IIB matrix model. 
The relevant supergravity solution to matrix model 
is supergravity solution with $b_{\mu\nu}$ field 
%\footnote{Here $b_{\mu\nu}$ is the second rank NS-NS 
%anti-symmetric tensor field. 
%This and $B_{\mu\nu}$ which appeared in the previous section 
%are different. } 
since the D-brane in matrix model is not pure one but 
one with 
$b_{\mu\nu}$ background. 
The alignment of D1-branes and D5-branes is same 
with the previous section. 
$x_{2}$, $x_{3}$, $x_{4}$, $x_{5}$, directions 
are compactified on a torus $T^{4}$.
This gives black string solution 
if $Q_{1}$ and $Q_{5}$ are large. 
We will work out in Euclidean space. 
The supergravity solution of D1-D5 system 
with the $b_{\mu\nu}$ fields which have non-zero components 
along all the directions parallel to the branes 
can be obtained likewise as in \cite{RSandBMM, MR}. 
The procedure is as follows. 
We apply T-duality transformation 
in the $x_{1}$ direction in a usual D1-D5 system. 
This gives D0-D4 bound states. 
Then we rotate the $(x_{0},x_{1})$-plane 
with a rotation angle $\theta$. 
Now we consider making the T-duality transformation 
in the $x_{1}$ (this is a rotated coordinate) 
direction. 
This gives the mixed boundary condition 
in the $x_{0}$, $x_{1}$ directions. 
In this T-duality transformation, 
$b_{01}= \tan\theta$ is induced 
along the $x_{0}$, $x_{1}$ directions. 
By applying the same procedure 
in the $x_{2}$, $x_{3}$ and $x_{4}$, $x_{5}$ directions, 
we obtain the following 
solution\footnote{The notation is referred to \cite{MR}.}: 
\begin{Eqnarray}
ds^{2}
= f_{1}^{-\frac{1}{2}}f_{5}^{-\frac{1}{2}} h_{1} 
  \left( dx_{0} ^{2} + dx_{1} ^{2} \right) 
  &+&f_{1}^{\frac{1}{2}}f_{5}^{\frac{1}{2}} 
  \left( dr^{2}+r^{2}d\Omega_{3}^{2} \right) \cr
 &+& f_{1}^{\frac{1}{2}} f_{5}^{-\frac{1}{2}}
  [h_{2}(dx_{2}^{2}+dx_{3}^{2}) 
    +h_{3}(dx_{4}^{2}+dx_{5}^{2})], \label{sugrad1d5}
\end{Eqnarray}
\begin{displaymath}
e^{2\phi}= g^{2} \frac{f_1}{f_5} h_{1}h_{2}h_{3}, 
\end{displaymath}
\begin{displaymath}
 f_{1,5} =1+\frac{\alpha^{\prime} R_{1,5}^{2}}{r}, 
\end{displaymath}
\begin{Eqnarray}
b_{01}&=& \tan \theta f_{1}^{-1}f_{5}^{-1} h_{1} \cr
b_{23}&=& \tan \theta f_{1}f_{5}^{-1} h_{2} \cr
b_{45}&=& \tan \theta 
  f_{1}f_{5}^{-1} h_{3} ,\nonumber
\end{Eqnarray}
where 
\begin{Eqnarray}
h_{1}^{-1}&=&f_{1}^{-1}f_{5}^{-1} \sin^{2} \theta +\cos ^{2} \theta, \cr
h_{2}^{-1}&=& h_{3}^{-1}= 
 f_{1}f_{5}^{-1} \sin^{2} \theta +\cos ^{2} \theta.  \nonumber
\end{Eqnarray}
This solution is considered in string frame.
$\theta$ is a parameter which represents 
noncommutativity. 

We consider the region which is very far from branes, 
that is, $r$ is very large. 
The spacetime become the flat space-time. 
We can ignore 
the $b_{\mu\nu}$ effect (i.e. noncommutativity) 
in this region.  
In such a region, 
the interactions are described as 
the interactions between diagonal blocks 
using the cluster property\cite{IKKT,AIIKKT} 
in matrix model. 
We consider the instanton and (anti-)instanton pair, 
each is in D5-brane, 
and the situation that their distance is very far. 
Instanton configuration in D5-brane 
is discussed at the last part in the previous section. 
We embed D1-D5 bound states 
in the left upper part of matrices and 
the other D1-D5 bound states in the left upper part of matrices. 
The exchange of gravitons, dilatons and axions 
can be seen 
similar to D3-D($-$1) case\cite{AIIKKT}. 
The interaction potential between instantons 
vanish because this system is BPS saturated. 
On the other hand, graviton exchange can be seen 
in the instanton and anti-instanton case\cite{AIIKKT}. 

It is conjectured by Maldacena\cite{Maldacena} 
that  two-dimensional CFT describing the Higgs branch 
of  D1-D5 system on $M^{4}$ is dual to 
IIB string on $AdS_{3} \times S^{3} \times M^{4}$. 
In the spirit of this relation, 
supergravity solution dual to noncommutative Yang-Mills 
is studied by Maldacena and Russo\cite{MR}, 
Hashimoto and Itzhaki\cite{HI}. 
Solution dual to two-dimensional gauge theory on D1-D5-branes 
with $b_{\mu\nu}$ field is 
obtained by taking the near horizon limit 
in the solution (\ref{sugrad1d5}), 
\begin{equation} 
\alpha^{\prime} \rightarrow 0, \hspace{0.5cm} 
\tan \theta = \frac{\tilde{b}}{\alpha^{\prime}}, 
\end{equation}
\begin{displaymath}
x_{0,1}=\frac{\alpha^{\prime}}{\tilde{b}}
    \tilde{x}_{0,1}, \hspace{0.5cm}
x_{2,3,4,5}=(\alpha^{\prime})^{\frac{1}{2}}\tilde{x}_{2,3,4,5}, 
\end{displaymath}
\begin{displaymath}
r=\alpha^{\prime}R^{2}u, \hspace{0.5cm} 
g=\alpha^{\prime} \tilde{g}, 
\end{displaymath}
where $\tilde{b}$, $\tilde{x}_{0,1}$, $\tilde{x}_{2,3,4,5}$, 
$u$ and $\tilde{g}$ stay fixed and $R^2=R_{1}R_{5}$. 

\begin{equation}
\frac{ds^{2}}{\alpha^{\prime}}
= R^{2} \left[ u^{2}\hat{h} \left( d\tilde{x}_{0} ^{2} 
+ d\tilde{x}_{1} ^{2} \right) 
    + \frac{dU^{2}}{U^{2}} + d\Omega _{3}^{2}   \right]
    + \frac{R_{1}}{R_{5}} d\tilde{x_n}d\tilde{x_n} , 
\end{equation}
\begin{displaymath}
\hat{h}=\frac{1}{1+a^{4}u^{4}}, 
 \hspace{0.5cm} a^{2}=\tilde{b}R^{2}, 
\end{displaymath}
\begin{displaymath}
e^{2\phi} = \hat{g}^{2} \hat{h}.
\end{displaymath}
where $\hat{g}=\tilde{g} \tilde{b}(R_{1}/R_{5})^{2}$ is 
the string coupling in the IR region. 
From the perspective of gauge theory, 
radial coordinate $u$ is interpreted as energy scale.
In the IR region, this metric represents 
$AdS_{3} \times S^{3} \times M^{4}$. 
The string coupling constant depends on the energy scale.
The string coupling decrease as the energy scale grows.
This is the same behavior 
with one in $AdS_{5} \times S^{5}$ 
which behaves like the twisted reduced model
\cite{AIIKKT}. 
The behavior of the dilaton is 
\begin{equation}
e^{\phi} \sim \hat{g} \frac{1}{a^{2}u^{2}}
  =\hat{g} \frac{1}{\tilde{b}R^{2} u^{2}}
\end{equation}
at large $u$.
This is in accordance with the fact that 
the relevant noncommutative Yang-Mills theory is two-dimensional
\cite{MRS}. 
\vspace{0.5cm}

In IIB matrix model, curved spacetime is thought to be given 
by assuming the following eigenvalue distribution\cite{IK}, 
\begin{equation}
\langle \rho(x) \rangle  \sim \sqrt{g} e^{-\phi(x)},  
\end{equation}
where 
\begin{equation}
\rho(x) =  \sum_{i} \delta^{(10)}(x-x^{i}). 
\end{equation}
In this case, we can calculate as follows , 
\begin{Eqnarray}
\sqrt{g} e^{-\phi(x)} &=& (\alpha^{\prime} R_1 R_5)^{\frac{5}{2}}
  \hat{g}^{-1} u\hat{h}^{\frac{1}{2}} \cr
&=& (\alpha^{\prime} R_1 R_5)^{\frac{5}{2}}\hat{g}^{-1}
\left( \frac{u^{2}}{1+ a^{4}u^{4}} \right) ^{\frac{1}{2}}, 
\end{Eqnarray} 
in only $AdS_3$ part.
This is invariant under $u \rightarrow 1/a^{2}u$ 
as expected. 
This eigenvalue distribution has a peak at $u=1/a$ 
while ordinary $AdS$ has a peak at $u=\infty$ (i.e. boundary).

%%%%%%%%%%%%%%%%%%%%%%%%%%%%%%%%%

\section{Discussions}

\hspace{0.4cm}
In this paper, we have considered D1-D5 system 
in IIB matrix model. 
Using the equivalence between twisted reduced model and 
noncommutative Yang-Mills, 
we studied the two-dimensional gauge theory 
on D1-D5 system. 
Twisted reduced model is obtained by the 
expansion around the noncommutative background. 
D1-D5 background is supersymmetric and one fourth of 
the supersymmetry survive. 
We discussed the spectrum of the fields, 
especially the fields which appear in the 
off-diagonal parts of matrices.
They correspond to the state which appear 
in  $(1,5)$ or $(5,1)$ open string.
We showed that only four degrees of freedom 
can be massless modes. This fact 
coincides with string theory result. 
D1-branes embedded in D5-branes are expressed 
as nontrivial solutions, instantons, 
in six-dimensional gauge theory on D5-branes.
We also considered the zero modes of the instanton configuration. 
The topological charge of the instanton is related to 
the existence of the chiral fermion zero modes 
by the index theorem. 
It was shown that there were $Q_{1}Q_{5}$ $SO(4)$  Weyl fermions 
as the zero modes.  
These degrees of freedom 
play an important role in the black hole physics.
%These chiral fermions correspond 
%to $\hat{\psi}_{3}$ in (\ref{fermion}).
We also considered the anti-instanton solution 
which is interpreted as and an anti-D1-brane 
embedded in a D5-brane. 
The anti-D1-brane background is constructed by 
replacing $\hat{x}_{1}$ 
with -$\hat{x}_{1}$ in (\ref{background}).
We can understand that anti-D1-brane and D5-brane 
system is unstable because this system is not supersymmetric. 
We can check it as follows.
We obtain 
$-C^{01}\sigma_{3}\Gamma^{01}\epsilon =1_{2\times 2}\xi$ 
instead of the first condition in (\ref{susy4}). 
It is impossible to find the spinors which satisfy 
this condition. 
However, an Anti-D1-brane can dissolve in a D5-brane 
and it is described as 
anti-instanton in the gauge theory on D5-brane. 
We have shown that this anti-instanton solution is supersymmetric, 
so this system can be stable as the bound state.    

In section 4, we discussed the supergravity solution 
which is dual to the gauge theory of the D1-D5 system in matrix model. 
In \cite{AIIKKT}, D-instantons were constructed as 
local excitations of gauge theory on D3-branes. 
It was also shown that D-instantons couple to gravity 
by considering the interactions between diagonal blocks. 
In this paper, we considered D1-brane solutions 
as local excitations of gauge theory on D5-branes. 
In matrix model, D1-brane is expressed by infinitely 
many D-instantons. 
Therefore the interactions between D1-branes and 
(anti-)D1-branes can be viewed as the interactions between 
infinitely many D-instantons and (anti-)D-instantons.

\vspace{1cm}
\begin{center}
{\bf Acknowledgments}
\end{center}
We would like to thank 
N.Ishibashi and 
the members of the KEK theory group 
for useful discussions. 
\vspace{0.5cm}

%%%%%%%%%%%%%%%%%%%%%%%%%%%%%%%%%%

\end{document}